\documentclass[aps,pra, preprint,groupedaddress,floatfix,longbibliography]{revtex4-1}
\usepackage{amsmath}
\usepackage{graphicx}
\usepackage[usenames,dvipsnames]{color}
\pdfoutput=1

\usepackage{color}
\usepackage{url}
\usepackage[colorlinks]{hyperref}
\hypersetup{%
        plainpages=true,
        breaklinks=true,
        hypertexnames=false,
        pageanchor=true,
        colorlinks=true,
        linkcolor={blue},
        citecolor={blue},
        urlcolor={black}, 
        anchorcolor={black}
      }

\renewcommand{\eqref}[1]{\mbox{Eq.~(\ref{#1})}}
\newcommand{\figref}[1]{\mbox{Fig.~\ref{#1}}}
\newcommand{\appref}[1]{\mbox{Appendix~\ref{#1}}}

\begin{document}


\title{Multiphoton Quantum Rabi Oscillations\\in Ultrastrong Cavity QED}

\author{Luigi Garziano$^{1,2}$}
\author{Roberto Stassi$^{1,2}$}
\author{Vincenzo Macr\`{i}$^1$}
\author{Anton Frisk Kockum$^2$}
\author{Salvatore Savasta$^{1,2}$}
\author{Franco Nori$^{2,3}$}
\affiliation{$^1$Dipartimento di Fisica e di Scienze della Terra, Universit\`{a} di Messina, I-98166 Messina, Italy,}
\affiliation{$^2$CEMS, RIKEN, Saitama 351-0198, Japan,}
\affiliation{$^3$Physics Department, The University of Michigan, Ann Arbor, Michigan 48109-1040, USA}



\begin{abstract}
When an atom is strongly coupled to a cavity, the two systems  can exchange a {\em single} photon through a coherent Rabi oscillation. This process enables precise quantum-state engineering and manipulation of atoms and photons in a cavity, which play a central role in quantum information and measurement.
Recently, a new regime of cavity QED has been reached experimentally where the strength of the interaction between light and artificial atoms (qubits) becomes comparable to the atomic transition frequency or the resonance frequency of the cavity mode. 
Here we show that this regime can strongly modify the concept of vacuum Rabi oscillations, enabling multiphoton exchanges between the qubit and the resonator.
We find that experimental state-of-the-art circuit-QED systems can undergo {\em two}- and {\em three}-photon vacuum Rabi oscillations. These anomalous Rabi oscillations can be exploited for the realization of efficient Fock-state sources of light and complex entangled states of qubits. 


\end{abstract}

\pacs{}

\maketitle


\section{Introduction}
Light-matter interaction in the strong-coupling regime is a coherent reversible process in which a
photon is absorbed and re-emitted by an electronic transition at a rate equal to the coupling
energy divided by the Planck constant \cite{Brune1996,Raimond2001}.
Reaching the light-matter strong-coupling regime has been a major focus of research in atomic physics and quantum optics for several decades and
has driven the field of cavity quantum electrodynamics (cQED) \cite{Thompson1992,Haroche2006}. 
The strong-coupling regime has been observed, both in the time and frequency domain, in a variety of systems \cite{Auffeves2013}, when an electronic transition is resonantly coupled to a cavity (optical resonator) and the coupling rate exceeds the rates of relaxation and decoherence of both the electronic transition and the field. 
Cavity QED effects with individual qubits have been intensively studied in solid state systems by replacing  natural atoms with artificial atoms, such as quantum dots \cite{Yoshie2004,Reithmaier2004} and Josephson circuits \cite{Blais2004, Wallraff2004, Chiorescu2004, You2011}.
The strong-coupling regime, when reached with a single qubit, enables a high degree of manipulation and control of quantum systems \cite{Haroche2013}.
For example, by exploiting the strong-coupling regime of cavity QED, the preparation and measurement of arbitrary quantum states in a completely controlled and deterministic manner has been achieved \cite{Deleglise2008, Hofheinz2009} and ``Schr\"{o}dinger's cat'' states of radiation have been prepared and reconstructed \cite{Haroche2013a, Vlastakis2013}.
Basic steps in quantum information processing, including the deterministic entanglement of atoms and the realization of quantum gates using atoms and photons as quantum bits have also been demonstrated \cite{Raimond2001, You2006, Kimble2008}.

Recently a new regime of cavity QED, where the coupling rate becomes an appreciable fraction of the unperturbed frequency
of the bare systems, has been experimentally reached in  a variety of solid state systems \cite{Gunter2009,Forn-Diaz2010,Niemczyk2010,Schwartz2011,Geiser2012,Scalari2012,Gambino2014,Goryachev2014,Maissen2014}. In this so called
ultrastrong-coupling (USC) regime, the routinely-invoked rotating wave approximation (RWA) is no longer applicable, and the antiresonant terms in the interaction Hamiltonian significantly change the standard cavity-QED scenarios (see, \emph{e.g.}, Refs.~\cite{Liberato2007, Ashhab2010, Cao2010, Cao2011, Ridolfo2012, Ridolfo2013, Stassi2013, Huang2014, Cacciola2014}).  
Although counter-rotating terms in principle exist in any real light-matter interaction Hamiltonian, their effects become prominent only in the USC limit \cite{Garziano2014}.
Usually, light-matter USC is reached by coupling the resonator with a large number of molecules or more generally electronic transitions. USC with a single qubit has been achieved only by using superconducting circuits based on Josephson junctions, which exhibit macroscopic quantum coherence and giant dipole moments as artificial atoms \cite{Niemczyk2010, Forn-Diaz2010,Baust2014}.

One interesting feature of the USC regime is that the number of excitations in the cavity-emitter system is no longer conserved, even in the absence of drives and dissipation. Measurements on a superconducting circuit QED system in the USC regime have shown clear evidence  of this feature \cite{Niemczyk2010}. Specifically, tuning the qubit transition frequency (by adjusting the external flux bias threading the qubit), and measuring the cavity transmission, an anticrossing arising from the coupling between states with a different number of excitations has been observed. In particular, the measurements evidence the coupling of  $|g,0,0,1 \rangle$ and  $|e,1,0,0 \rangle$, where the kets indicate the states of the qubit and of the first three resonator modes.
Very recently it has been shown \cite{Law2015} that,  when the frequency of the
cavity field is near one-third of the atomic transition frequency, there exists a resonant three-photon coupling via intermediate states connected by counter-rotating processes.

The resonant quantum Rabi oscillations, occurring when the atom and the cavity mode can exchange one excitation quantum  in a reversible way, 
play a key role in the manipulation of atomic and field states for quantum information processing \cite{Haroche2013}.
Here, we show that a system consisting of a single qubit ultrastrongly coupled to a resonator can exhibit anomalous vacuum Rabi oscillations where {\em two} or {\em three} photons are jointly emitted by the qubit into the resonator and re-absorbed by the qubit in a reversible and coherent process. We focus on the case of a flux qubit coupled to a coplanar-waveguide resonator, a system where the USC regime with a single artificial atom has been demonstrated \cite{Niemczyk2010}. We find that this effect can be observed at coupling rates of the same order of those already reached in these systems \cite{Niemczyk2010,Baust2014}. 

\section{Dissipation and Photodetection in the USC regime}
In order to demonstrate multiphoton quantum Rabi oscillations in USC cavity QED, we calculate  the time evolution of the mean output photon fluxes and higher-order normally-ordered photon correlations. 

It has been shown that, in the USC regime, the usual normally-ordered correlation functions fail to describe the output photon emission rate and photon statistics. Clear evidence of this is that the standard input-output relations  predict, even for a vacuum input and the system in the ground state, a finite output photon flux proportional to the average number of cavity photons \cite{DeLiberato2009,Beaudoin2011,Ridolfo2012}, {\em i.e.}, $\langle \hat A_{\rm out}^{-}(t) \hat A_{\rm out}^{+}(t)\rangle\propto \langle \hat a^{\dagger}(t) \hat a(t)\rangle$, where $\hat A_{\rm out}^{+}(t)$ and $\hat A_{\rm out}^{-}(t)$ are the positive- and negative-frequency components of the output field operator, while  $\hat a$ and $\hat a^\dag$ are the destruction and creation operators for cavity photons. 
A solution to this problem has been proposed in Ref.\ \cite{Ridolfo2012}. Considering for the sake of simplicity a single-mode resonator, it is possible to derive the correct output photon emission rate and correlation functions by expressing the cavity electric-field operator $\hat X = \hat a + \hat a^\dag$ in the atom-cavity dressed basis. Once the cavity electric-field operator has been expressed in the dressed basis, it has to be decomposed in its positive- and negative-frequency components $\hat X^{+}$ and $\hat X^{-}$  \cite{Ridolfo2012}. Expanding the $\hat X$ operator in terms of the energy eigenstates $|j\rangle$ (with $\hbar \omega_j $ the corresponding eigenvalues) of the system Hamiltonian $\hat H$, one finds the relations
\begin{equation}
\label{5}
\hat X^{+}=\sum_{j,k>j}X_{jk}|j\rangle \langle k| \;; \; \hat X^{-}=(\hat X^{+})^{\dagger},\end{equation} 
where $X_{jk}\equiv \langle j| (\hat a^{\dagger}+\hat a)|k\rangle$ and the states  are labeled such that $\omega_{k} > \omega_{j}$ for $k>j$. The resulting positive frequency output operator can be expressed as
\begin{equation}
\hat A_{\rm out}^{+}(t) = \hat A_{\rm in}^{-}(t) - \sqrt{\kappa}  C_0 \hat X^+, 
\label{in-out}
\end{equation}
where $\kappa$ is the loss rate of the resonator due to the coupling to the external in-out modes and $C_0$ is a constant proportional to the zero-point fluctuation amplitude of the resonator \cite{Garziano2013}.
Two aspects of these results are noteworthy: first of all, we note that in the USC regime, one correctly obtains $\hat X^{+}|0 \rangle=0$ for the system in its ground state $|0 \rangle$ in contrast to $\hat a|0\rangle\neq 0$. Moreover, we notice that the positive-frequency component of $\hat X$ is not simply proportional to the photon annihilation operator $\hat a$. As a consequence, for arbitrary degrees of light-matter interaction, the output photon flux emitted by a resonator can be expressed as $\Phi_{\rm out}= \kappa \langle \hat X^{-} \hat X^{+}\rangle$. Similarly, the output delayed coincidence rate is proportional to the two-photon correlation function $\langle \hat X^{-}(t) \hat X^{-}(t + \tau) \hat X^{+}(t+ \tau) \hat X^{+}(t)\rangle$. In quantum optics, it is well known that  the signal directly emitted from the qubit is proportional to $\langle \hat \sigma_+ \hat \sigma_- \rangle$. In circuit QED systems, this emission can be detected by coupling the qubit to an additional microwave antenna \cite{Hofheinz2009}. Indeed, in  the USC regime the qubit emission rate becomes proportional to the qubit mean excitation number $\langle \hat C^- \hat C^+ \rangle$, where $\hat C^\pm$ are the qubit positive and negative frequency operators, defined as $\hat C^{+}=\sum_{j,k>j}C_{jk}|j\rangle \langle k|$ \;and \; $\hat C^{-}=(\hat C^{+})^{\dagger}$, with $C_{jk}\equiv \langle j|(\hat \sigma_{-}+ \hat \sigma_{+})|k\rangle$.

In order to properly describe the system dynamics, including dissipation and decoherence effects, the coupling to the environment needs to be considered.
We adopt the master-equation approach. However, in the USC regime the description offered by the standard quantum-optical master equation breaks down \cite{Agarwal2013}.
Following Ref.~\cite{Breuer2002,Beaudoin2011}, we write the system operators in the system-bath interaction Hamiltonian in a basis formed by the eigenstates of $\hat H$. We consider $T=0$ temperature reservoirs (the generalization to $T\neq0$ reservoirs is straightforward). By applying the standard Markov approximation and tracing out the reservoir degrees of freedom, we arrive at the master equation for the density matrix operator $\hat \rho(t)$,
\begin{equation}\label{ME}
\dot{\hat\rho}= \frac{i}{\hbar} [\hat \rho(t),\hat H] + \mathcal{L}_{\rm damp}\hat\rho(t)+ \mathcal{L}_{\phi}\hat\rho(t)\, ,
\end{equation}
where $\mathcal{L}_{\rm damp} \hat \rho(t)=\sum_{j,k>j}(\Gamma_{\kappa}^{jk}+\Gamma_{\gamma}^{jk}) \mathcal{D}[|j\rangle\langle k|]\hat \rho(t)$ with $\mathcal{D}[\hat O]\hat\rho=\frac{1}{2}(2 \hat O\hat\rho \hat O^{\dagger}-\hat \rho \hat O^{\dagger}\hat O-\hat O^{\dagger}\hat O \hat \rho)$, describes dissipation effects arising from the resonator and  qubit reservoirs. These cause  transitions  between eigenstates at rates
\begin{align}
\label{rate1}
\Gamma_{\kappa}^{jk} = \kappa |\langle j| \hat X |k \rangle|^2, \\ 
\Gamma_{\gamma}^{jk} = \gamma |\langle j| \hat \sigma_x |k \rangle|^2, 
\end{align}
where $\kappa$ and $\gamma$ are decay rates, here assumed to be spectrally constant, induced by the resonator and qubit reservoirs.
Pure dephasing effects affecting the qubit are described in Eq.\ (\ref{ME}) by the last term $\mathcal{L}_{\phi}\hat\rho(t) = {\mathcal D} [\sum_j \Phi_{j} |j \rangle \langle j|]\hat\rho(t)$, where $\Phi_{j} = \sqrt{\gamma_\phi/2} \langle j| \hat \sigma_z | j \rangle$, and $\gamma_\phi$ is the pure dephasing rate. Note that only the most relevant diagonal contributions have been included.

\section{Results}

Here we study a flux qubit coupled to a coplanar resonator in the USC regime. 
For suitable junctions, the qubit potential landscape is a double-well potential, where the two minima correspond to
states with clockwise and anticlockwise persistent currents $\pm I_{\rm p}$ \cite{Chiorescu2004, You2011}. When the flux offset $\delta \Phi_x \equiv \Phi_{\rm ext} - \Phi_0/2 = 0$, where $\Phi_{\rm ext}$ is the external flux threading the qubit and $\Phi_0$ is the flux quantum,  the lowest two energy states are separated by an energy gap $\Delta$. In the qubit eigenbasis, the qubit Hamiltonian reads $\hat H_{\rm q}  =  \hbar \omega_{\rm q} \hat \sigma_z /2 $, where  $\hbar \omega_{\rm q} = \sqrt{\Delta^2 + (2I_{\rm p} \delta \Phi_x)^2}$ is the qubit transition frequency, which can be adjusted by an external flux
bias. We note that  the two-level approximation is well justified because of the large anharmonicity of this superconducting artificial atom. The resonator modes are described as harmonic oscillators, $\hat H_{m} = \hbar \omega^{\rm r}_m \hat a_m^\dag \hat a_m$, where $\omega^{\rm r}_m$ is the resonance frequency, $m$ is the resonator-mode index and $\hat a^\dag_m$ ($\hat a_m$) is the bosonic creation (annihilation) operator for the $m$-th resonator mode. We will consider $\lambda/2$ and $\lambda/4$ resonators. Then, the quantum circuit can be described by the following extended Rabi Hamiltonian
\begin{equation}
\hat H = \hat H_{\rm q} + \sum_m [ \hat H_m + \hbar g_m \hat X_m
 (\cos \theta\,  \hat \sigma_x + \sin \theta\, \hat \sigma_z )].
\label{H}
\end{equation}

Here, $\hat X_m = \hat a_m + \hat a_m^\dag $, $\hat \sigma_{x,z} $ denote Pauli operators, $g_m$ is the coupling rate of the
qubit to the $m$-th cavity mode and the flux dependence is encoded
in $\cos \theta = \Delta/ \omega_{\rm q}$. The operator $\hat \sigma_{x}$  is conveniently
expressed as the sum of the qubit raising ($\hat \sigma_{+}$) and lowering ($\hat \sigma_{-}$)
operators, which in the Heisenberg picture and for $g_m=0$ oscillate as $\exp{(i \omega_{\rm q} t)}$ (negative frequency) and $\exp{(-i \omega_{\rm q} t)}$ (positive frequency) respectively. Thus, in contrast to the Jaynes-Cummings (JC) model \cite{Jaynes1963}, the Hamiltonian in Eq.~(\ref{H}) explicitly contains counter-rotating
terms of the form $\hat \sigma_+ \hat a_m^\dag$, $\hat \sigma_- \hat a_m$, $\hat \sigma_z \hat a_m^\dag$, and $\hat \sigma_z \hat a_m$. 
Considering only one resonator mode and a flux offset $\delta \Phi_x =0$, the Hamiltonian in Eq.~(\ref{H}) reduces to the standard Rabi Hamiltonian.
\begin{figure}[ht]
  \includegraphics[height= 110 mm]{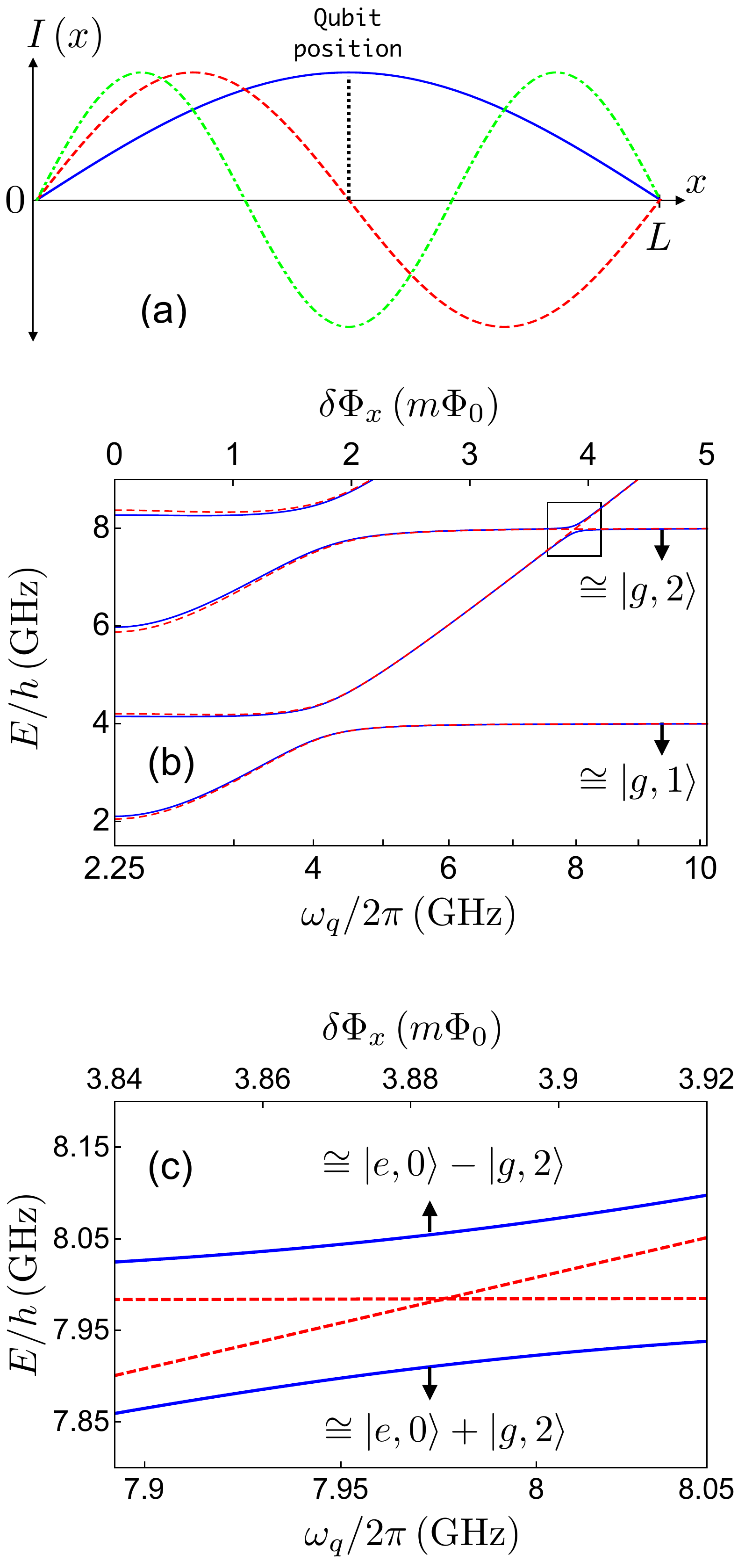}
  \caption{(Color online) (a) Sketch of the distribution of the first three resonator modes $(m=1,2,3)$ of a transmission-line $\lambda/2$ resonator. The resonance frequency of the first mode (blue solid curve) is set to $\omega^{\rm r}_1/2\pi= 4$ GHz. The qubit is positioned at the center of the resonator, so that it does not interact with the $m=2$ (red dashed curve) and is off-resonance to the $m=3$ (green dotted-dashed curve) mode. The bare resonance frequencies of the second and third modes are $\omega^{\rm r}_2= 2 \omega^{\rm r}_1 $ and $\omega^{\rm r}_3= 3 \omega^{\rm r}_1 $, respectively. Qubit parameters are $\Delta/h=2.25$ GHz and $2I_{\rm p}=630$ nA.
(b) Frequency differences $\omega_{i,0} = \omega_{i} - \omega_{0}$ for the lowest energy dressed  states as a function of the qubit transition frequency $\omega_{\rm q}$ (which can be tuned by changing the external flux bias $\delta \Phi_x$) for the JC model (red dashed curves) and the extended Rabi Hamiltonian (blue solid curves). We consider a normalized coupling rate $g_1/\omega^{\rm r}_1=0.15$ between the qubit and the resonator. In both cases the ground state level is not displayed.
(c) Avoided level crossing (blue solid curves) resulting from the coupling between the states $|e,0 \rangle$ and $|g,2 \rangle$ due to the presence of counter-rotating terms in the system Hamiltonian. The energy splitting reaches its minimum at $\omega_{\rm q}/2\pi \approx 7.97 \;\rm{GHz} \approx 2 (\omega^{\rm r}_1/2\pi)$. The anticrossing is not present in the JC model (red dashed lines), since it arises from the coherent coupling between states with a different number of excitations. \label{fig:2PhEnergyLevels}}
\end{figure}

\subsection{Two-photon quantum Rabi oscillations}
We first consider the case of a flux qubit coupled to a  $\lambda/2$ superconducting transmission-line resonator with resonance frequencies $\omega^{\rm r}_m=m\pi c/L$, where $L$ is the resonator length. We use the qubit parameters $\Delta/h=2.25$ GHz and $2I_{\rm p}=630$ nA, as in Ref.~\cite{Niemczyk2010}. We are interested in the situation where the qubit 
transition energy is approximately twice that of the fundamental resonator mode: $\omega_{\rm q} \approx 2 \omega^{\rm r}_1$. We consider the qubit to be positioned at the center of the resonator, so that it does not interact with the resonator mode $m=2$ (see \figref{fig:2PhEnergyLevels}a). 
The other resonator modes are much higher in energy, detuned with respect to the qubit transition frequency by an amount significantly larger than the coupling rate: $(\omega^{\rm r}_m - \omega_{\rm q}) \approx (m-2) \omega^{\rm r}_1$, providing only moderate energy shifts for the coupling rates $g_m/ \omega^{\rm r}_m \lesssim 0.2$ considered here. We will thus only take into account the interaction of the qubit with the fundamental resonator mode. We diagonalize numerically the Hamiltonian from Eq.~(\ref{H}) and indicate the resulting energy eigenvalues and eigenstates as $\hbar \omega_{i}$ and  $| i\rangle$  with $i = 0, 1, \dots$, choosing the labeling of the states  such that $\omega_{k} > \omega_{j}$ for $k>j$.     

Figure~\ref{fig:2PhEnergyLevels}b shows the frequency differences $\omega_{i,0} = \omega_{i} - \omega_{0}$ for the lowest energy states as a function of the qubit transition frequency, which can be tuned by changing the external flux bias $\delta \Phi_x$. The red dashed curves correspond to calculations obtained neglecting all the counter-rotating terms (JC model). We observe a spectrum with two large-splitting anticrossings around $\omega_{\rm q} \approx \omega_1^r$ which appear both in the dashed and the continuous curves. In the JC picture, they correspond to the resonant coupling between states with the same number of excitations. The lowest energy avoided crossing results from the coherent coupling of the states $|e,0 \rangle$ and $|g, 1 \rangle$, where $g\, (e)$ indicates the ground (excited) state of the qubit and the second entry in the kets represents the photon number. When the splitting reaches its minimum, the resulting system eigenstates are 
\begin{equation}\label{JC state}
 \frac{1}{\sqrt{2}}(|e,0 \rangle  \pm |g, 1 \rangle)\,.
\end{equation}
The higher-energy large avoided crossing in the plot corresponds to the second rung of the JC ladder, arising from the coupling of $|e,1 \rangle$ and $|g, 2 \rangle$. Only small quantitative deviations between the eigenenergies in the JC and in the extended Rabi model can be observed.

When the counter-rotating terms are taken into account, the states $|i \rangle$ are no longer eigenstates of the total number of the excitation operator $\hat N_{\rm exc} =\hat a^\dag a + \hat \sigma_+ \hat \sigma_-$. For example, the system ground state can be expressed as a superposition  $|  0 \rangle = \sum_n c^{0}_{gn} |g, n\rangle +c^{0}_{en} |e, n\rangle$ of bare states also involving nonzero excitations. Of course, when the normalized coupling $g_1/\omega^{\rm r}_1 \ll 1$, only the coefficient $c^{0}_{g0}$
 is significantly different from zero. Moreover, for $\theta= 0$ parity is conserved \cite{Liu2005,Braak2011} and only states with an even number of excitations contribute to  $|0 \rangle$.
The non-conservation of the total excitation number also affects the excited dressed states $|j \rangle = \sum_n c^{j}_{gn} |g, n\rangle +c^{j}_{en} |e, n\rangle$. As a consequence, the dressed states $|1 \rangle$ and $|2 \rangle$ at the minimum splitting do not correspond to the simple JC picture of Eq.~(\ref{JC state}).

The continuous line levels in \figref{fig:2PhEnergyLevels}b also display a  smaller amplitude avoided crossing when $\omega_{\rm q} \approx 2 \omega^{\rm r}_1$. Observing that just outside this avoided-crossing region one level remains flat as a function of the flux offset $\delta \Phi_x$ with energy $\omega \approx 2 \omega^{\rm r}_1$ while the other shows a linear behavior with $\omega_{\rm q}$, the splitting originates from the hybridization of the states $|e,0 \rangle$ and $|g, 2 \rangle$. This avoided crossing behavior is better shown in \figref{fig:2PhEnergyLevels}c, and the resulting states are well approximated by the states $\frac{1}{\sqrt{2}}(|e,0 \rangle  \pm |g, 2 \rangle)$.
This splitting is not present in the RWA, where the coherent coupling between states with a different number of excitations is not allowed, nor does it occur with the standard Rabi Hamiltonian ($\theta =0$).

Following the procedure described in Ref.~\cite{Law2015}, such a two-photon coupling between the bare states $|e,0 \rangle$ and $|g, 2 \rangle$ can be analytically described by an effective Hamiltonian (see Appendix). As displayed  in \figref{fig:2PhotonDiagram}, the coupling between $|e,0 \rangle$ and $|g, 2 \rangle$ can only occur via the intermediate states $|g, 1 \rangle$ and $|e, 1 \rangle$. Indeed, if the system is initially prepared in the state $|e,0 \rangle$, two different processes can occur: either (i) the counter-rotating term $\hat a_1^\dag \hat \sigma_z $ enables a virtual transition  $|e,0 \rangle \rightarrow |e,1 \rangle$, and then the term $\hat a_1^\dag \hat \sigma_{-}$  leads to the final transition to the state $|g,2 \rangle$; or (ii) the term $\hat a_1^\dag \hat \sigma_{-}$ enables the transition $|e,0 \rangle \rightarrow |g,1 \rangle$ which is followed by  the virtual transition $|g,1 \rangle \to |g,2 \rangle$ induced by the term $\hat a_1^\dag \hat \sigma_z$.
\begin{figure}[t!]
\includegraphics[height= 55 mm]{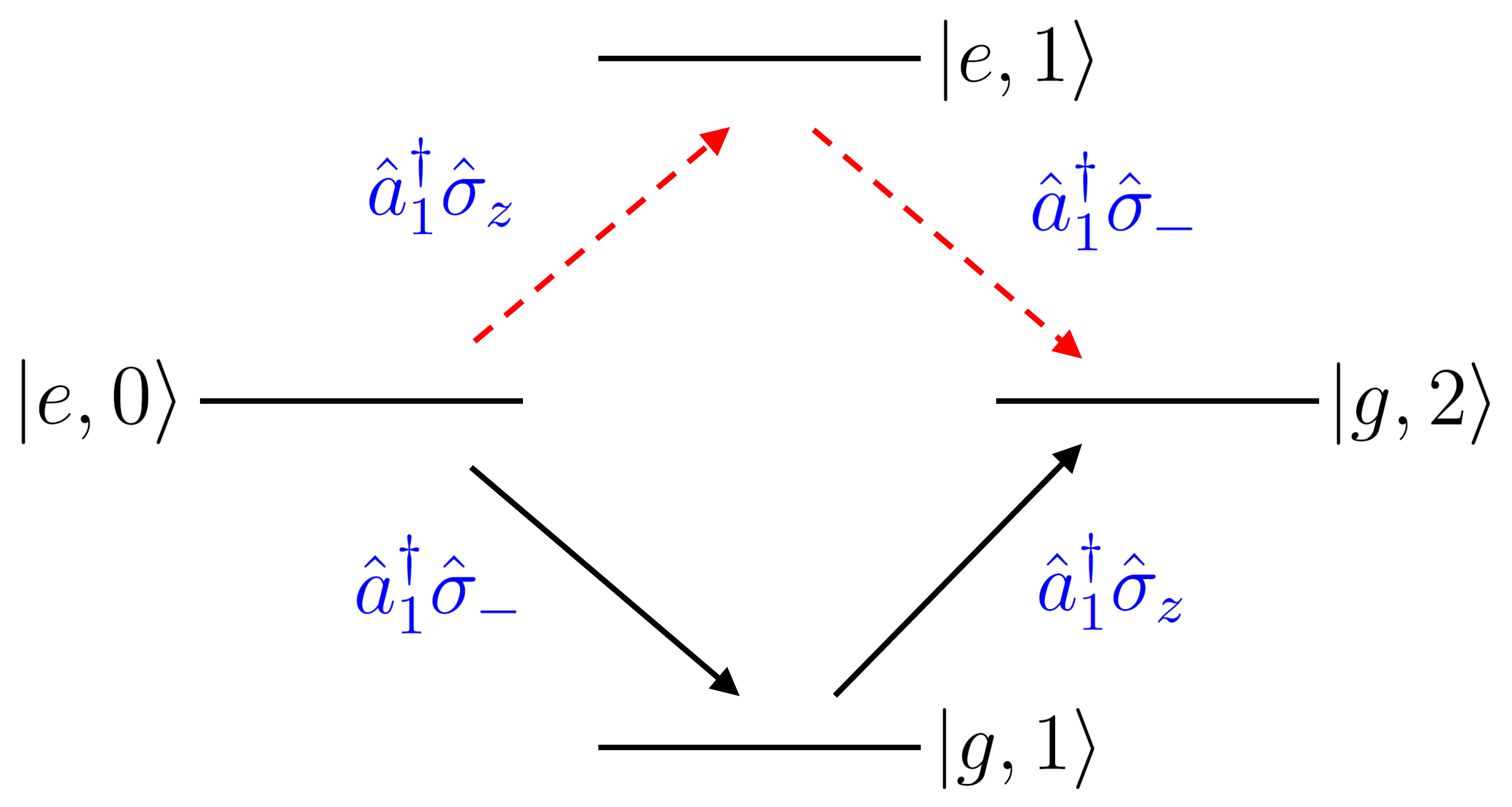}
\caption{(Color online) Coupling between the bare states $|e,0\rangle$ and $|g,2 \rangle$ via the intermediate states $|g,1\rangle$ and $|e,1 \rangle$. The red dashed arrows and the solid black arrows indicate, respectively, the two different processes describing the two-photon resonant coupling between $|e,0\rangle$ and $|g,2 \rangle$. \label{fig:2PhotonDiagram}}
\end{figure} 

In order to obtain an analytical description of the effective coupling, we first reduce the  extended Rabi Hamiltonian to the truncated Hilbert space composed of the bare states $|e,0 \rangle$, $|g,1 \rangle$, $|e,1 \rangle$, and $|g,2 \rangle$.
 The matrix form of the reduced Hamiltonian becomes
\\
\begin{equation}\label{Hred}
\frac{\hat H_{\rm r}}{\hbar}=\begin{pmatrix}
\frac{\omega_{\rm q}}{2} & g_1 \cos \theta & g_1 \sin \theta & 0\\ 
 g_1 \cos \theta & \omega^{\rm r}_1-\frac{\omega_{\rm q}}{2} & 0 & -\sqrt{2} g_1 \sin \theta\\ 
  g_1 \sin \theta & 0  & \omega^{\rm r}_1+\frac{\omega_{\rm q}}{2} & \sqrt{2} g_1 \cos \theta\\ 
 0 &  -\sqrt{2} g_1 \sin \theta & \sqrt{2} g_1\cos \theta & 2\omega^{\rm r}_1-\frac{\omega_{\rm q}}{2}
 \end{pmatrix},
\end{equation}
where the order of columns and rows is $|e,0 \rangle$, $|g,1 \rangle$, $|e,1 \rangle$, and $|g,2 \rangle$.

Near the two-photon resonance when $\omega_{\rm q} \approx 2 \omega^{\rm r}_1$, the intermediate states $|g,1 \rangle$ and $|e,1 \rangle$ can be adiabatically eliminated (see \appref{app:Derivation}), leading to the effective Hamiltonian
\\
\begin{equation}\label{Heff}
\begin{split}
\hat H_{\rm eff}&=  \bigg(\frac{\omega_{\rm q}}{2}+\frac{2g_1^2}{\omega_{\rm q}} \cos(2\theta)\bigg)|e,0\rangle\langle e,0|\\
&+\bigg(2\omega^{\rm r}_1-\frac{\omega_{\rm q}}{2}-\frac{4g_1^2}{\omega_{\rm q}} \cos(2\theta) \bigg)|g,2\rangle\langle g,2|\\
&-\Omega_{\rm eff}^{\rm (2ph)}(|e,0\rangle\langle g,2|+|g,2\rangle\langle e,0|),
\end{split}
\end{equation}
which describes the effective two-photon coupling between $|e,0 \rangle$ and $|g,2 \rangle$, with an effective two-photon Rabi frequency
\begin{equation}\label{omegaeff}
\Omega_{\rm eff}^{\rm (2ph)}\equiv   \frac{2 \sqrt{2 }g_1^2\sin(2\theta)}{\omega_{\rm q}}.
\end{equation}

A key theoretical issue of the USC regime is the distinction between bare (unobservable) excitations and physical particles that can be detected.
For example, when the counter-rotating terms are relevant, the mean photon number in the system ground state is different from zero: $\langle 0|\hat a^{\dagger} \hat a|0 \rangle\neq 0$.
However, these photons are actually virtual since they do not correspond to real particles that can be detected in a photon-counting experiment.
According to this analysis, the presence of an $n$-photon contribution in a specific eigenstate of the system does not imply that the system can emit $n$ photons when prepared in this state.
\begin{figure}[ht]
  \includegraphics[height= 90 mm]{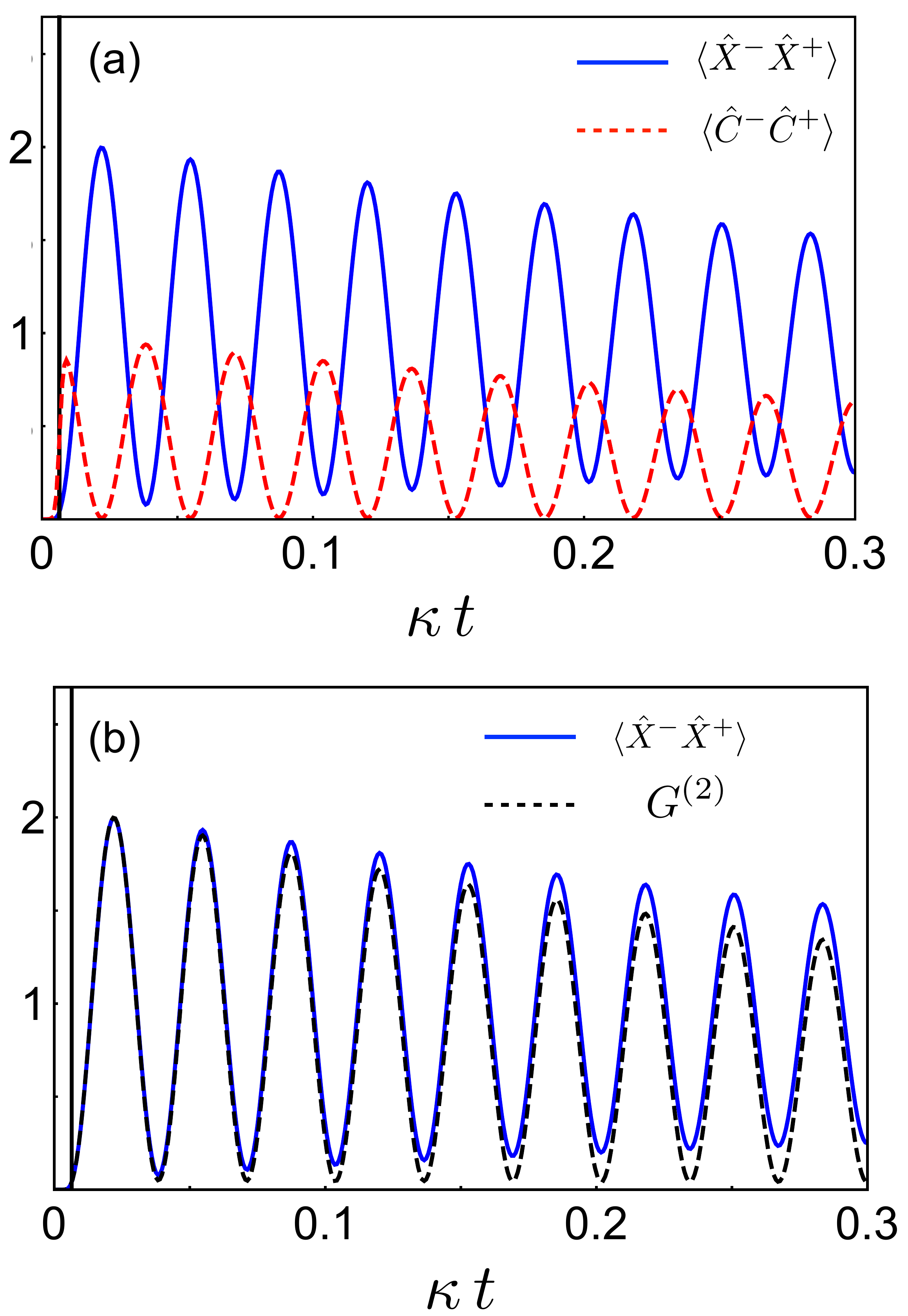}
  \caption{(Color online) (a) Temporal evolution of the cavity mean photon number $\langle \hat X^{-} \hat X^{+}\rangle$ (blue solid curve) and the qubit mean excitation number $\langle \hat C^{-} \hat C^{+}\rangle$ (red dashed curve) after the arrival of a $\pi$-like Gaussian pulse exciting the qubit (the black vertical line shows the wavepacket peak arrival time). The amplitude and the central frequency of the pulse are $A/\omega^{\rm r}_1= 8.7 \times10^{-2}$ and $\omega = (\omega_{3,0} + \omega_{2,0})/2$, respectively. After the arrival of the pulse, the system undergoes vacuum Rabi oscillations showing the reversible excitation exchange of two photons  between the qubit and the resonator.
(b) Time evolution of the zero-delay two-photon correlation function $G^{(2)}(t)$ (dashed black curve) together with the intracavity photon number $\langle \hat X^{-} \hat X^{+}\rangle$ (blue solid curve). At early times they almost coincide. 
This perfect two-photon correlation is a signature that photons are actually emitted in pairs.
Resonator and qubit damping rates are  $\kappa/\omega^{\rm r}_1=1.8 \times10^{-4}$ and $\gamma/\omega^{\rm r}_1=1.8 \times10^{-4}$, respectively. \label{fig:2PhCorrelationFunction}}
\end{figure}
In order to fully understand and characterize this avoided crossing not present in the RWA, a more quantitative analysis is required.
In the following, we therefore calculate the output signals and correlations which can be measured in a photodetection experiment.

\begin{figure}[ht]
  \includegraphics[height= 140 mm]{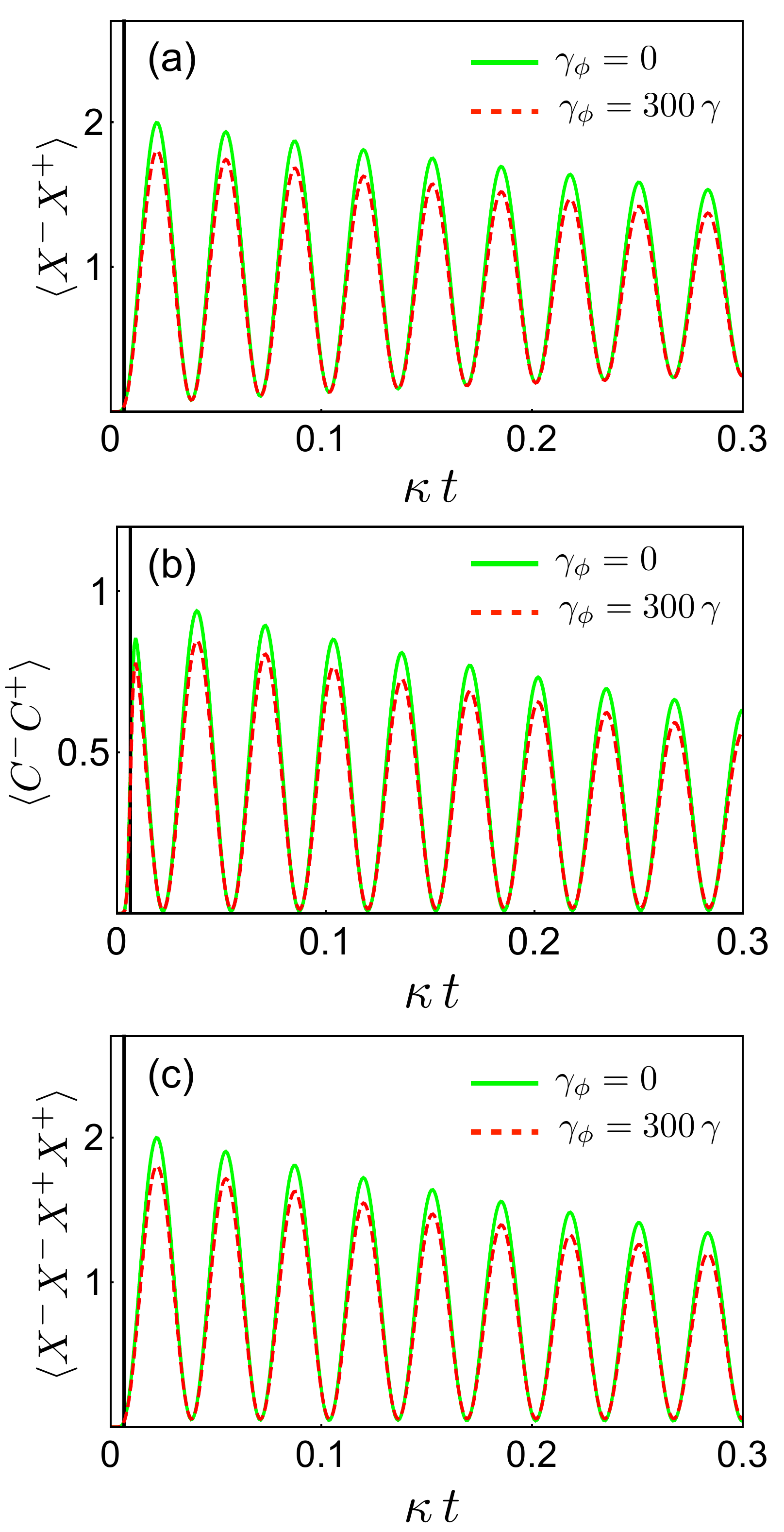}
  \caption{\label{Fig4}(Color online) Effects of strong pure dephasing (dashed lines) on the dynamics of the mean photon number (a), the qubit effective population (b), and the two-photon correlation function (c). Calculations have been performed with the same parameters as in \figref{fig:2PhCorrelationFunction} with the addition of a pure dephasing rate $\gamma_\phi = 300 \;\gamma$. Solid lines display numerical results obtained in the absence of pure dephasing $(\gamma_\phi =0)$. It can be observed that the physics of multiphoton vacuum Rabi oscillations is not significantly altered by the effects of pure dephasing. \label{fig:2PhDephasing}}    
\end{figure}
In order to probe the anomalous avoided crossing shown in Figs.~\ref{fig:2PhEnergyLevels}b and \ref{fig:2PhEnergyLevels}c, we consider the case where the qubit is directly excited via a microwave antenna by an optical Gaussian pulse. The corresponding driving Hamiltonian is
\begin{equation}\label{pulse}
	\hat H_{\rm d} = {\cal E}(t) \cos (\omega t) \hat \sigma_x\,,   
\end{equation}
where ${\cal E}(t) = A \exp{[-(t-t_0)^2/(2 \tau^2)]}/(\tau \sqrt{2 \pi})$. Here, $A$ and $\tau$ are the amplitude and the standard deviation of the Gaussian pulse, respectively. We consider the zero-detuning case by choosing the flux offset $\delta \Phi_x$ corresponding to the qubit frequency $\omega_{\rm q}/2 \pi \simeq 7.97$ GHz, where the splitting in \figref{fig:2PhEnergyLevels}b is at its minimum. The central frequency of the pulse has been chosen to be in the middle of the two split transition energies: $\omega = (\omega_{3,0} + \omega_{2,0})/2$.

The cavity output photon flux and the photon flux emitted by the qubit directly coupled to a microwave antenna are proportional to $\langle \hat X^- \hat X^+ \rangle$ and $\langle \hat C^- \hat C^+ \rangle$, respectively. Figure~\ref{fig:2PhCorrelationFunction}a displays the dynamics of these two quantities after the arrival of a $\pi$-like pulse exciting the qubit described by the Hamiltonian (\ref{pulse}). Calculations in \figref{fig:2PhCorrelationFunction} have been carried out in the absence of pure dephasing ($\gamma_\phi =0$). Results for $\gamma_\phi \neq 0$ are shown in \figref{fig:2PhDephasing}. Vacuum Rabi oscillations showing the reversible excitation exchange between the qubit and the resonator are clearly visible in \figref{fig:2PhCorrelationFunction}a. 
The pulse time-width is not much narrower than the Rabi period, so the qubit excitation is partially transferred to the cavity during the pulse arrival. Therefore, the first peak of the qubit mean excitation number (\figref{fig:2PhCorrelationFunction}a) is slightly lower than the second one.
We observe that the mean intracavity physical photon number at its first maximum is very close to two. This is a first hint that when the qubit is in the ground state the resonator mode acquires two photons. However, the output measured signals are proportional and not equal to  $\langle \hat X^- \hat X^+ \rangle$, so from this kind of measurements it is not possible to certify that the qubit and the resonator are actually exchanging two quanta.

Figure~\ref{fig:2PhCorrelationFunction}b displays the time evolution of the zero-delay two-photon correlation function $G^{(2)}(t)=\langle \hat X^{-}(t) \hat X^{-}(t ) \hat X^{+}(t) \hat X^{+}(t)\rangle$ (dashed black line) together with the intracavity photon number $\langle \hat X^{-} \hat X^{+}\rangle$ (blue continuous curve) for comparison. At early times they almost coincide. This is a signature of perfect two-photon correlation: the probability of the system to emit one photon is equal to the probability to emit a photon pair. During the system evolution, higher values of the local minima of $\langle \hat X^{-} \hat X^{+}\rangle$ are observed due to the decay from the two-photon state to the one-photon state (out of resonance with respect to the qubit) caused by the photon escape from the resonator. However, the two-photon correlation function $\langle \hat X^{-}(t) \hat X^{-}(t ) \hat X^{+}(t) \hat X^{+}(t)\rangle$ goes almost to zero every time the qubit is maximally excited.  This different behavior in \figref{fig:2PhCorrelationFunction}b indicates that the qubit does not absorb single photons but only photon pairs. The period of a complete population oscillation is $2 \pi /\Omega_{\rm eff}^{\rm (2ph)}$, where $2\Omega_{\rm eff}^{\rm (2ph)}$ is the minimum splitting in \figref{fig:2PhEnergyLevels}c.  

Ordinary quantum vacuum oscillations have already been demonstrated in circuit-QED systems ({\em e.g.}, \cite{Chiorescu2004,Johansson2006}). The dynamics observed in \figref{fig:2PhCorrelationFunction} can also be obtained by first preparing the qubit in its excited state by employing a $\pi$ pulse. Then the $\pi$ pulse can be followed by a shift pulse which brings the qubit into resonance with the resonator for the desired duration in order to observe the coupled dynamics as in Ref.\ \cite{Johansson2006}.
If the shift pulse has a duration $\delta t = 2 \pi / \Omega_{\rm eff}^{\rm (2ph)}$, the Fock state $n =2$ is directly generated. After the switch-off of the shift pulse, the qubit is out of resonance with the resonator and the Fock state can escape from the cavity through an input-output port and be detected. Hence two-photon Rabi oscillations can be exploited for fast and efficient generation of two-photon states.

The influence of strong pure dephasing effects is shown in \figref{fig:2PhDephasing}. Calculations have been performed with the same parameters used in \figref{fig:2PhCorrelationFunction} with the addition of a pure dephasing rate $\gamma_\phi = 300 \gamma$. Figure~\ref{fig:2PhDephasing} compares the dynamics of the mean photon number (a), the qubit effective population (b), and the two-photon correlation (c) in the absence (continuous curves) and in the presence (dashed curves) of pure dephasing. The figure shows that strong pure dephasing does not significantly alter the physics of multiphoton vacuum Rabi oscillations and the main effect of dephasing is to make the excitation pulse less effective.

\subsection{Three-photon quantum Rabi oscillations}
\begin{figure}[ht]
  \includegraphics[height= 110 mm]{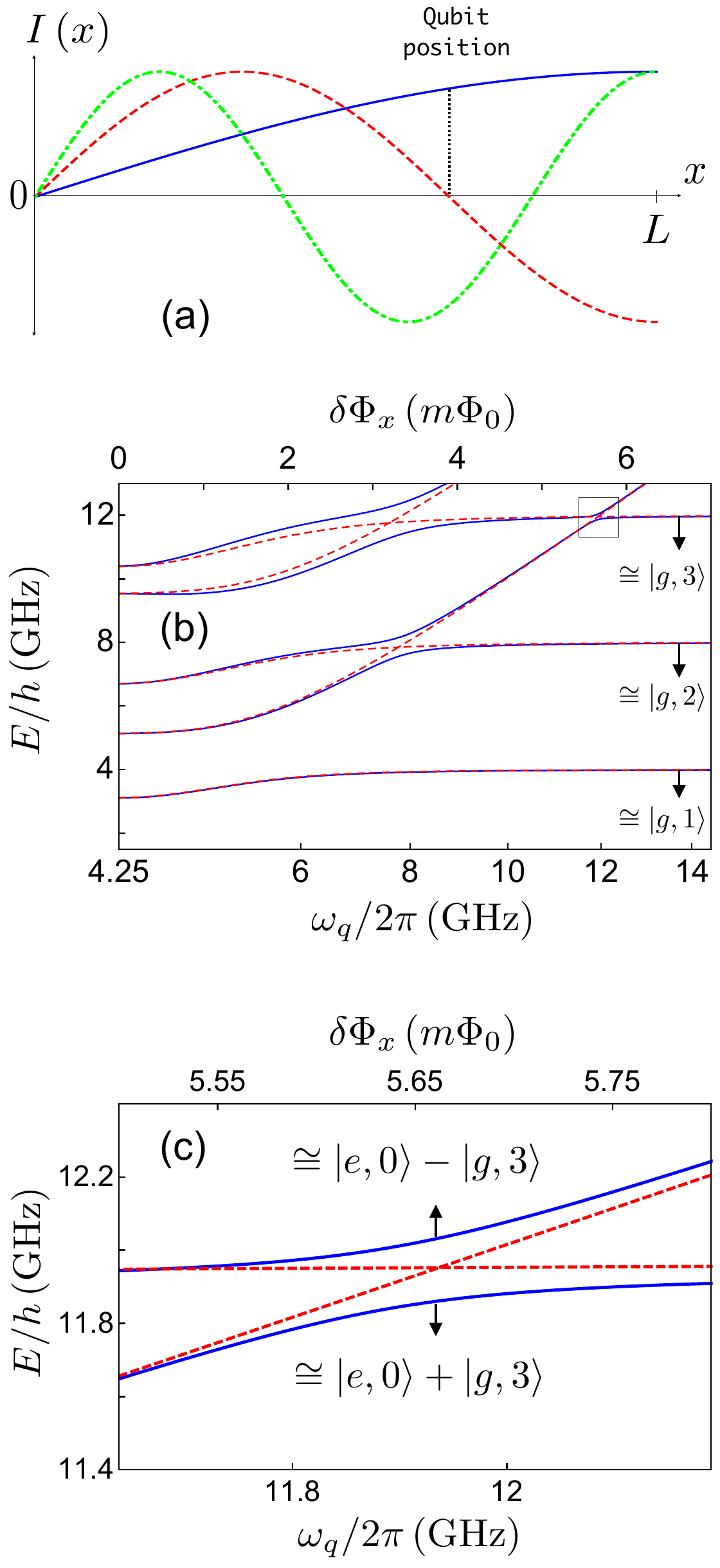}
  \caption{(Color online) (a) Sketch of the distribution of the first three resonator modes $(m=1,2,3)$ of a transmission-line $\lambda/4$ resonator. The resonance frequency of the first mode (blue solid curve) is $\omega^{\rm r}_1= 4$ GHz. The qubit is positioned so that it does not interact with the $m=2$  mode (red dashed curve)  with resonance frequency $\omega^{\rm r}_2 = 3 \omega^{\rm r}_1$. The mode $m=3$ (green dotted-dashed curve)  has resonance frequency $\omega^{\rm r}_3  = 5 \omega^{\rm r}_1$, which is much larger than $\omega_{\rm q}$ so that  only the interaction of the qubit with the fundamental resonator mode can be considered. Qubit parameters are $\Delta/h=4.25$ GHz and $2I_{\rm p}=630$ nA.
 (b) Frequency differences $\omega_{i,0} = \omega_{i} - \omega_{0}$ for the lowest dressed energy states as a function of the qubit transition frequency $\omega_{\rm q}$ (which can be tuned by changing the external flux bias $\delta \Phi_x$) for the JC model (red dashed curves) and the extended Rabi Hamiltonian (blue solid curves) explicitly containing counter-rotating terms. We consider a normalized coupling rate $g_1/\omega^{\rm r}_1=0.25$ between the qubit and the resonator.
The spectrum shows two large-splitting anticrossings, which appear only in the continuous curves, plus a smaller avoided crossing which is magnified in (c).
(c) Three-photon vacuum Rabi splitting (blue solid curves) resulting from the coupling between the states $|e,0 \rangle$ and $|g,3 \rangle$ due to the presence of counter-rotating terms in the system Hamiltonian. The energy splitting reaches its minimum at $\omega_{\rm q}/2\pi\approx 11.89 \;\rm{GHz} \approx 3(\omega^{\rm r}_1/2\pi)$. The anticrossing is not present in the JC model (red dashed curves), since it arises from the coherent coupling between states with a different number of excitations. \label{fig:3PhEnergyLevels}}
\end{figure}

Very recently, it has been shown \cite{Law2015} that the strong coupling of a single qubit with three photons can be achieved in the USC regime when the frequency of the cavity field is near one-third of the atomic transition frequency.
In this case, parity-symmetry breaking is not required and this effect can occur also at $\theta = 0$. Hence, it could be observed also in systems like natural atoms or molecules displaying parity symmetry. 
One possible problem with this configuration is that the qubit can also interact resonantly with the one-photon state of the $m=3$ mode of the resonator. In this case the qubit would interact with both one- and three-photon states.
We show that the undesired one-photon resonant coupling of the qubit with a resonator mode can be avoided by considering a $\lambda /4$ resonator, whose resonance frequencies are $\omega^{\rm r}_m=(2m-1)\pi c/2L$, with $m=1, 2, 3 \dots$ (see \figref{fig:3PhEnergyLevels}a).
 As shown in \figref{fig:3PhEnergyLevels}a, the qubit is positioned so that it does not interact with the mode $m=2$ with resonance frequency $\omega^{\rm r}_2  = 3 \omega^{\rm r}_1$. The qubit parameters are $\Delta/h=4.25$ GHz and $2I_{\rm p}=630$ nA. 
In the present case we are interested in the situation where the qubit transition energy is approximately three times that of the fundamental resonator mode: $\omega_{\rm q} \approx 3 \omega^{\rm r}_1$. The mode $m = 3$ has resonance frequency $\omega^{\rm r}_3  = 5 \omega^{\rm r}_1$, which is much larger than $\omega_{\rm q}$. Hence also in this case, we can consider the interaction of the qubit with only the fundamental resonator mode. 

Figure~\ref{fig:3PhEnergyLevels}b displays the frequency differences $\omega_{i,0}$ for the lowest energy states as a function of the qubit transition frequency. The red dashed curves
 corresponds to calculations obtained neglecting all the counter-rotating terms (JC model). We observe a spectrum with two large-splitting anticrossings which appear only in the continuous curves plus a smaller avoided crossing magnified in \figref{fig:3PhEnergyLevels}c.
The lowest energy splitting corresponds to a two-photon vacuum Rabi splitting. When it reaches its minimum (at $\omega_{\rm q} \approx 2 \omega^{\rm r}_1 $), the corresponding hybridized states are analogous to those whose dynamics has been described in \figref{fig:2PhCorrelationFunction}. They can be approximately expressed as
\begin{equation}\label{states1}
 \frac{1}{\sqrt{2}}(|e,0 \rangle  \pm |g, 2 \rangle)\, .
\end{equation}
The second avoided crossing at higher energy corresponds to the second rung of the two-photon Rabi ladder and the corresponding approximated hybridized states (at the minimum splitting) are
\begin{equation}\label{states2}
 \frac{1}{\sqrt{2}}(|e,1 \rangle  \pm |g, 3 \rangle)\,.
\end{equation}
The third smaller splitting, occurring at $\omega_{\rm q} \approx 3 \omega^{\rm r}_1 $, corresponds to a three-photon vacuum Rabi splitting. Here, a single qubit is resonantly coupled with a three-photon state, resulting, at the minimum splitting, in the approximated eigenstates
\begin{equation}
\label{states3}
 \frac{1}{\sqrt{2}}(|e,0 \rangle  \pm |g, 3 \rangle)\, .
\end{equation}

\begin{figure}[ht]
  \includegraphics[height= 100 mm]{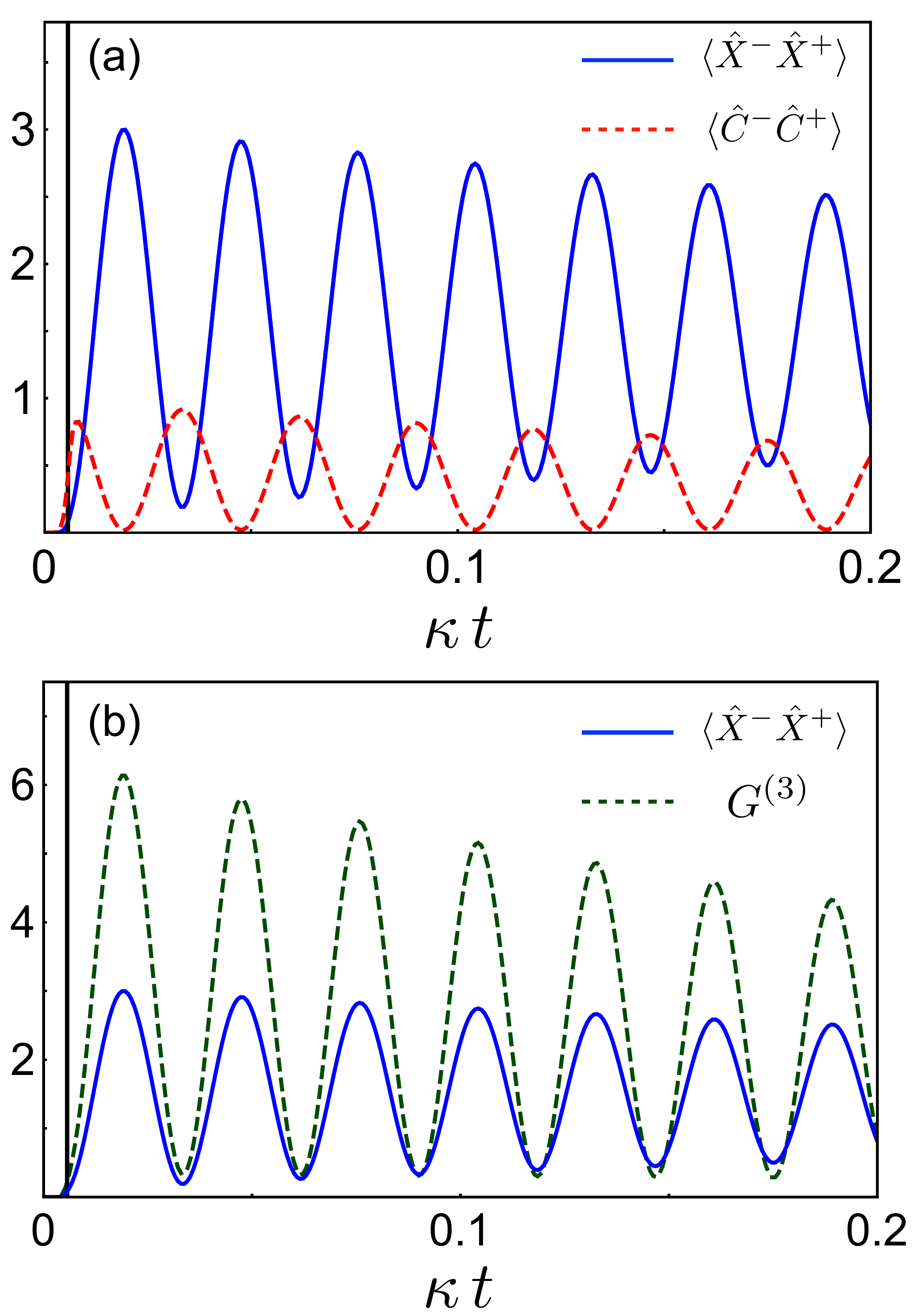}
  \caption{(Color online) (a) Temporal evolution of the cavity mean photon number $\langle \hat X^{-} \hat X^{+}\rangle$ (blue solid curve) and the qubit mean excitation number $\langle \hat C^{-} \hat C^{+}\rangle$  (red dashed curve) after the arrival of a $\pi$-like Gaussian pulse exciting the qubit (the black vertical curve shows the wavepacket center time). The amplitude and the central frequency of the pulse are $A/\omega^{\rm r}_1= 9.4 \times10^{-2}$ and $\omega = (\omega_{4,0} + \omega_{3,0})/2$, respectively. After the arrival of the pulse, the system undergoes vacuum Rabi oscillations showing the reversible excitation exchange between the qubit and the resonator. The fact that the mean intracavity physical photon number at its first maximum is very close to three is a first signature that, when the qubit is in its ground state, the resonator mode is in a three-photon state.
	(b) Time evolution of the zero-delay three-photon function $G^{(3)}(t)$ (dashed green curve together with the intracavity photon number $\langle \hat X^{-} \hat X^{+}\rangle$ (solid blue curve). The first peak value of the three-photon correlation function is approximately two times higher than that of the mean photon number, a signature of an almost-perfect three-photon correlation. Parameters for resonator and qubit losses are the same as in \figref{fig:2PhCorrelationFunction}. \label{fig:3PhCorrelationFunction} 
	}
\end{figure}
Figure~\ref{fig:3PhCorrelationFunction} displays the system dynamics after the arrival of a $\pi$-like pulse  exciting the qubit described by the Hamiltonian (\ref{pulse}).
Specifically, \figref{fig:3PhCorrelationFunction}a shows the time evolution of $\langle \hat X^- \hat X^+ \rangle$ and $\langle \hat C^- \hat C^+ \rangle$.
Calculations have been carried out in the absence of pure dephasing ($\gamma_\phi =0$). Vacuum Rabi oscillations showing the reversible excitation exchange between the qubit and the resonator are clearly visible. We observe that the mean intracavity physical photon number at its first maximum is very close to three. This is a first hint that when the qubit is in the ground state the resonator mode is in a three-photon state. The period of a complete population oscillation is $2 \pi / \Omega_{\rm eff}^{\rm (3ph)}$, where $2\Omega_{\rm eff}^{\rm (3ph)}$ is the minimum splitting in \figref{fig:3PhEnergyLevels}c.

Figure~\ref{fig:3PhCorrelationFunction}b displays the time evolution of the zero-delay three-photon correlation function $G^{(3)}(t)= \langle \hat X^{-}(t) \hat X^{-}(t) \hat X^{-}(t ) \hat X^{+}(t) \hat X^{+}(t) \hat X^{+}(t)\rangle$ together with the intracavity photon number $\langle \hat X^{-} \hat X^{+}\rangle$ for comparison. At early times the peak values of $G^{(3)}(t)$ are approximately two times higher than those of the mean photon number $\langle \hat X^{-}(t) \hat X^{+}(t)\rangle$. This is a specific feature of three-photon Fock states and indicates an almost perfect three-photon correlation. We observe that $G^{(3)}(t)$ at early times reach a peak value slightly beyond 6. This indicates that the system has a nonnegligible probability to emit more than three photons. This is confirmed by the presence of a non-zero four-photon correlation function.
Analyzing the different transitions contributing to $G^{(3)}(t)$, we can attribute this effect to additional low-frequency transition $|4 \rangle \to |3 \rangle$. These transitions between Rabi-split states occurs when parity symmetry is broken, \cite{Ridolfo2011} and in this case
produces a 4-photon cascade:  $|4 \rangle \to |3 \rangle \to |2 \rangle \to |1 \rangle \to |0\rangle$.
This small contribution cannot be observed if its low frequency is outside the frequency-detection window. 
Analogously to the two-photon case (see \figref{fig:2PhDephasing}), pure dephasing does not significantly affect the dynamics of multiphoton quantum Rabi oscillations (plot not shown).

\subsection{Generation of entangled GHZ states}
Standard vacuum Rabi oscillations have been exploited for the realization of atom-atom entanglement. 
Here we show that the multiphoton Rabi oscillations can be directly applied to the deterministic realization of more complex entangled states. As a first application we discuss the deterministic realization of multi-atom Greenberger-Horne-Zeilinger (GHZ) states \cite{Greenberger1990} by using only one resonator. The GHZ states lead  to striking violations of local realism and are an important resource for quantum information processing \cite{Bose1998}, quantum cryptography  \cite{Cleve1999} and  error correction protocols \cite{DiVincenzo1996}. Superconducting circuits have been used to study GHZ states (see, {\em e.g.}, Ref.~\cite{Wei2006,Yang2015}).

Consider a resonator coupled to qubit 1 in the USC regime where two-photon vacuum Rabi oscillations can occur. The resonator also interacts in the strong (not ultrastrong) coupling regime with two additional qubits (2 and 3). Although the coupling rates between the resonator and the qubits are fixed, the qubit-resonator interaction can be switched on and off by adjusting the qubit frequencies \cite{Hofheinz2009}. The protocol is simple and consists of three steps, one for each qubit.
We start by exciting the ultrastrongly-coupled  qubit with a $\pi$-pulse. Then, by changing the flux offset (at time $t=0$), we drive it into resonance with the two-photon state of the resonator (\figref{fig:2PhEnergyLevels}c). The system state at time $t$ is $|\psi \rangle = \cos{(\Omega_{\rm eff}^{\rm (2ph)} t)} |e,g,g,0 \rangle + \sin (\Omega_{\rm eff}^{\rm (2ph)} t) |g,g,g,2 \rangle$. We let the qubit interact for a $\pi/2$ Rabi rotation so that the resulting state is  $(|e,g,g,0 \rangle + |g,g,g,2 \rangle) / \sqrt{2}$. We then drive the qubit back out of resonance, stopping the Rabi rotation. The second step consists of driving qubit 2 into resonance with the 1-photon state of the first  resonator mode for a $\pi$ rotation time  so that the resulting state is: $(|e,g,g,0 \rangle - |g,e,g,1 \rangle)/ \sqrt{2}$. For the third step, we similarly drive the third qubit into resonance with the first  resonator mode for a $\pi$ rotation. The resulting state is: $(|e,g,g,0 \rangle + |g,e,e,0 \rangle)/ \sqrt{2}$. At this point, the photon state can be factored out, leaving us with a three-qubit  GHZ-like entangled state. A more conventional GHZ state can be obtained by sending a further $\pi$ pulse to the first qubit, so that the resulting final state is  $(|g,g,g,0 \rangle + |e,e,e,0 \rangle)/ \sqrt{2}$.
This procedure can be easily generalized to four qubits or more. In general, if $n$-photon Rabi oscillations are achieved, then $n+1$ qubit GHZ states can be produced. We notice that this protocol does not need the initial synthesis of photonic or atomic superposition states \cite{Cirac1994,Zheng2001}.

\section{Conclusion}

We have investigated vacuum Rabi oscillations in the USC regime.
According to the Jaynes-Cummings model, the qubit and the resonator can exchange a single excitation quantum through a coherent Rabi oscillation process. Such Rabi oscillations play a key role in the manipulation of atomic and field states for quantum information processing \cite{Haroche2013}.
Our theoretical predictions show clear evidence for physics beyond the Jaynes-Cummings model and extend the concept of quantum Rabi oscillations.
We find that multiphoton reversible exchanges between an individual qubit and a resonator can be observed in the USC regime.
Specifically, we have shown that experimental state-of-the-art circuit-QED systems can undergo two- and three-photon vacuum Rabi oscillations. Still increasing the coupling rate, a higher number of photons can be exchanged with the qubit during a single Rabi oscillation.
These anomalous Rabi oscillations can be exploited for the realization of efficient Fock-state sources of light, and for the implementation of novel protocols for the control and manipulation of atomic and field states.

\section*{Acknowledgements}

\noindent We thank O. Di Stefano and Chui-Ping Yang for useful discussions. This work is partially supported by the RIKEN iTHES Project, the MURI Center for Dynamic Magneto-Optics via the AFOSR award number FA9550-14-1-0040,
the IMPACT program of JST, a Grant-in-Aid for Scientific Research (A), and  from the MPNS
COST Action MP1403 Nanoscale Quantum Optics.

\appendix

\section{Analytical derivation of the two photon-qubit effective Hamiltonian}
\label{app:Derivation}

In this Appendix, we derive the analytical expression for the effective Hamiltonian in Eq.~(\ref{Heff}), describing the two-photon coupling between the states $|e,0\rangle$ and $|g,2\rangle$. We start from the reduced Hamiltonian in Eq.~(\ref{Hred}) and then move to the rotating frame with frequency $\omega_{\rm q}/2$, obtaining the transformed reduced Hamiltonian
\begin{equation}\tag{A1}\label{Hrmod}
\frac{\hat H'_{\rm r}}{\hbar}=\begin{pmatrix}
0 & g_1 \sin \theta & g_1 \cos \theta & 0\\ 
 g_1 \sin \theta & \omega^{\rm r}_1 & 0 & \sqrt{2} g_1 \cos \theta\\ 
  g_1 \cos \theta & 0  & \omega^{\rm r}_1-\omega_{\rm q} & -\sqrt{2} g_1 \sin \theta\\ 
 0 &  \sqrt{2} g_1 \cos \theta & -\sqrt{2} g_1\sin \theta & 2\omega^{\rm r}_1-\omega_{\rm q}\, .
 \end{pmatrix}.
\end{equation}
Now, the order of columns and rows is $|e,0 \rangle$, $|e,1 \rangle$, $|g,1 \rangle$, and $|g,2 \rangle$.
After the transformation, an arbitrary state of the system in this truncated Hilbert space  can be denoted as $(c1,c2,c3,c4)^T$ and the Schr\"{o}dinger equation with Hamiltonian $\hat H'_{\rm r}$ gives
\begin{align}\tag{A2}
i\dot{c}_1&=(g_1 \sin\theta) c_2+(g_1 \cos\theta)c_3 \\
\tag{A3} i\dot{c}_2&= \omega^{\rm r}_1 c_2+ (g_1 \sin\theta) c_1+(\sqrt{2}g_1 \cos\theta)c_4 \\
\tag{A4} i\dot{c}_3&= (\omega^{\rm r}_1-\omega_{\rm q}) c_3+ (g_1 \cos\theta) c_1-(\sqrt{2}g_1 \sin\theta)c_4\\
\tag{A5}i\dot{c}_4&= (2\omega^{\rm r}_1-\omega_{\rm q}) c_4+ (\sqrt{2}g_1 \cos\theta)c_2-(\sqrt{2}g_1 \sin\theta)c_3\, .
\end{align}
For $g_1/\omega^{\rm r}_1\ll 1$, the adiabatic elimination in Eqs.~(A3) and (A4) can be applied \cite{Law2015} and the coefficients $c_2$ and $c_3$ can be approximated as 
\begin{align}\tag{A6}
c_2&\approx-\frac{g_1}{\omega^{\rm r}_1}(\sin \theta c_1+\sqrt{2}\cos \theta c_4) \\
\tag{A7}c_3&\approx -\frac{g_1}{(\omega^{\rm r}_1-\omega_{\rm q})}(\cos \theta c_1-\sqrt{2}\sin \theta c_4)\, .
\end{align}
The coupled equations for $c_1$ and $c_4$ are obtained substituting these results in Eqs.~(A2) and (A5):
\begin{equation}\tag{A8}
i\dot{c}_1 \approx \frac{g^2_1 (\omega_{\rm q} \sin^2 (2\theta)-\omega^{\rm r}_1)}{\omega^{\rm r}_1(\omega^{\rm r}_1-\omega_{\rm q})}c_1+\frac{\sqrt{2} g^2_1 \omega_{\rm q}\sin (2\theta)}{2\omega^{\rm r}_1(\omega^{\rm r}_1-\omega_{\rm q})}c_4\, ,
\end{equation}
\begin{equation}\tag{A9}
i\dot{c}_4 \approx \frac{-2g^2_1 (\omega^{\rm r}_1-\omega_{\rm q} \cos^2 \theta )+\omega^{\rm r}_1[2(\omega^{\rm r}_1)^2-3\omega^{\rm r}_1\omega_{\rm q}+\omega_{\rm q}^2]}{\omega^{\rm r}_1(\omega^{\rm r}_1-\omega_{\rm q})}c_4+\frac{\sqrt{2} g^2_1 \omega_{\rm q}\sin (2\theta)}{2\omega^{\rm r}_1(\omega^{\rm r}_1-\omega_{\rm q})}c_1\, .
\end{equation}
\begin{figure}[ht]
	\includegraphics[height= 55 mm]{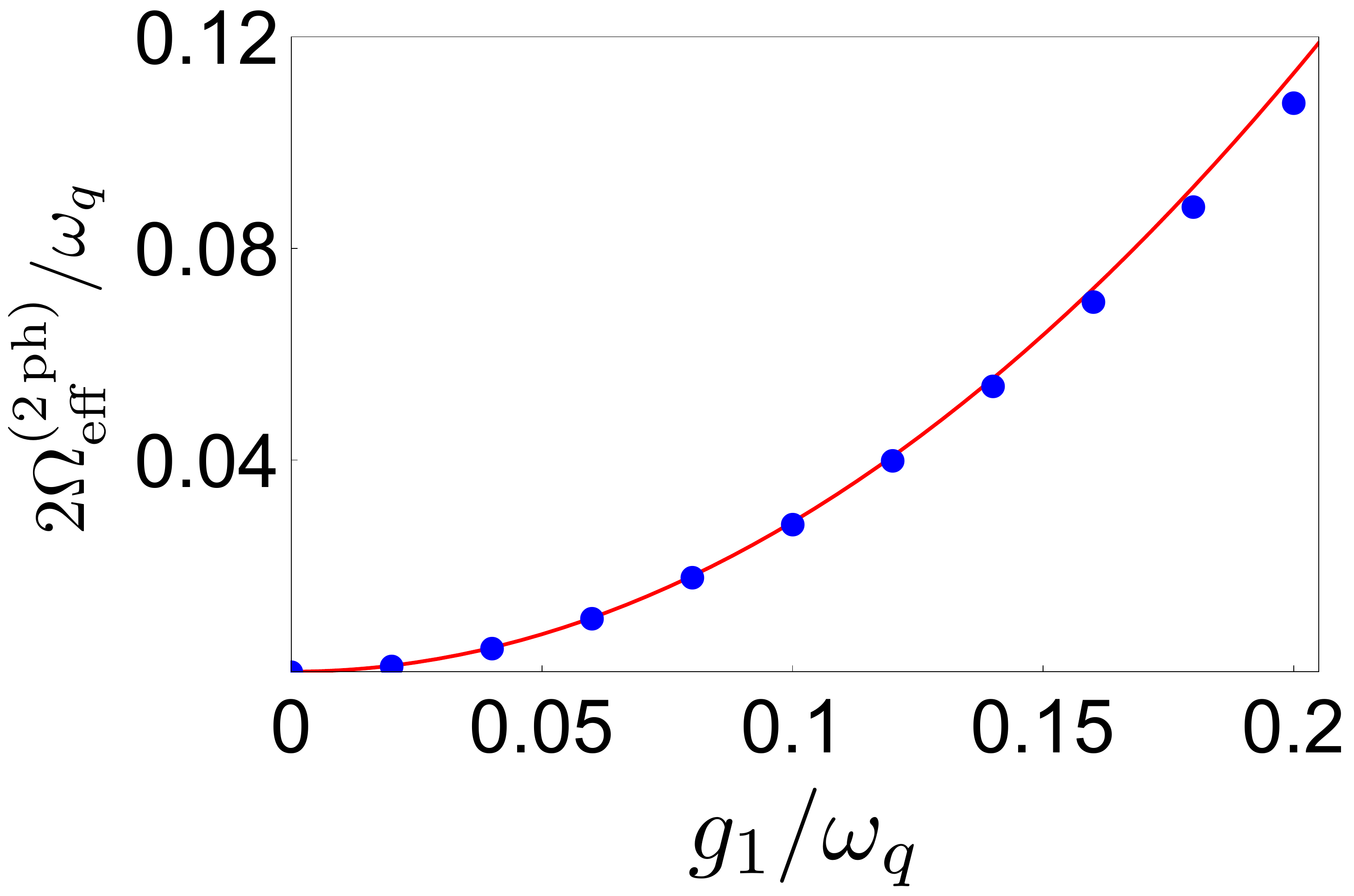}
	\caption{(Color online) Comparison between the minimum energy splitting $2\Omega_{\rm eff}^{\rm (2ph)}/\omega_{\rm q}$ obtained analytically (red solid line) and numerically (blue points) as a function of $g_1/\omega_{\rm q}$ for $\theta=\pi/4$. \label{fig:ComparingOmega}}
\end{figure}
Considering the near-resonant case $\omega^{\rm r}_1\approx \omega_{\rm q}/2$, transforming back to the laboratory frame and keeping only the $g^2_1$ dependence terms in the diagonal elements, the effective Hamiltonian
in Eq.~(\ref{Heff}) is obtained. According to the effective Hamiltonian (\ref{Heff}), the ratio of the minimum splitting at the avoided crossing (see \figref{fig:2PhEnergyLevels}c) to the qubit frequency $\omega_{\rm q}$ is given by
\begin{equation}\tag{A10}
\frac{2\Omega_{\rm eff}^{\rm (2ph)}}{\omega_{\rm q}}= 4 \sqrt{2}\sin(2\theta)\bigg(\frac{g_1}{\omega_{\rm q}}\bigg)^2.
\end{equation}
A comparison between analytical and numerical results for the minimum energy splitting $2\Omega_{\rm eff}^{\rm (2ph)}/\omega_{\rm q}$ is shown in \figref{fig:ComparingOmega} as a function of $g_1/\omega_{\rm q}$.


\bibliography{Riken}

\end{document}